\documentclass[proceedings]{JHEP3}

\PrHEP{PrHEP hep2001}                   
\conference{International Europhysics Conference on HEP}                

\usepackage{epsfig}                   

\newcommand {\xgo} {\mbox{$x_{\gamma}^{\rm{obs}}$}}
\newcommand{\dspm}       {\mbox{$D^{\ast \pm}$}}
\newcommand{\dsspm}      {\mbox{$D_s^{\pm}$}}
\def\ptr{p_T^{\rm rel}}

\title{Heavy Flavor in Photoproduction at HERA}

\author{\speaker{Leonid Gladilin}\thanks{On behalf of the H1 and ZEUS Collaborations.}\\  
        DESY, Notkestr. 85, 22603 Hamburg, Germany\\                                          
        E-mail: \email{gladilin@mail.desy.de}}                       

\abstract{Recent results on charm and beauty photoproduction
at HERA are discussed.
The perturbative QCD calculations are generally smaller than
the measured cross sections, particularly
in the forward (proton) direction.
The study of charm dijet photoproduction
is consistent with a significant contribution of charm excitation
processes.
}

\begin{document}

  \section{Introduction}

During the first phase of operation (1992--2000),
HERA collided electrons and positrons with energy
$26.7-27.6\,$GeV and protons
with energy $820-920\,$GeV
yielding a center-of-mass energy of $296-318\,$GeV.
In this period,
extensive measurements of heavy flavor photoproduction
were made by the H1 and ZEUS
collaborations~\cite{h1_dstar,dstar,dsubs,h1_bb,zeus_bb}.
Some recent charm and beauty photoproduction results
are discussed in this paper.

In photoproduction processes at HERA, a quasi-real photon
with virtuality $Q^2 \sim 0$ is emitted by the incoming
electron and interacts with the proton.
At leading order (LO) in QCD,
two types of processes are responsible for the production 
of heavy quarks:                       
the direct photon processes,
where the photon participates as a point-like particle,
and the resolved photon processes, where 
the photon acts as a source of partons.
The dominant direct photon process is photon-gluon fusion
where the photon fuses with a gluon from the incoming proton.
In resolved photon processes, a parton from the photon
scatters off a parton from the proton.
Charm and beauty quarks present in the parton distributions of the photon, as
well as of the proton, lead to processes
like $cg~\to~cg$ and $bg~\to~bg$,
which are called heavy flavor excitation processes.     
In next-to-leading order (NLO) QCD, only the sum of direct and resolved
processes is unambiguously defined. 

  \section{Charm photoproduction}

Differential cross sections
for \dspm\ and \dsspm\ photoproduction
in $p^D_\perp$ and $\eta^D$
were measured in the kinematic range
$Q^2 < 1$\,GeV$^2$, $130 < W < 280$\,GeV,
$3<p_{\perp}^{D}<12$\,GeV and
$|\eta^{D}|<1.5 $~\cite{dstar,dsubs}.
Here
$W$ is the $\gamma p$ center-of-mass energy, and
$p_{\perp}^{D}$ and $\eta^{D}$ are
the $D$-meson transverse momentum and
pseudorapidity, respectively.
The pseudorapidity $\eta$ is defined as $-\ln(\tan \frac{\theta}{2})$,
where the polar angle $\theta$ is measured with respect to the proton beam direction.

\FIGURE{
\epsfig{figure=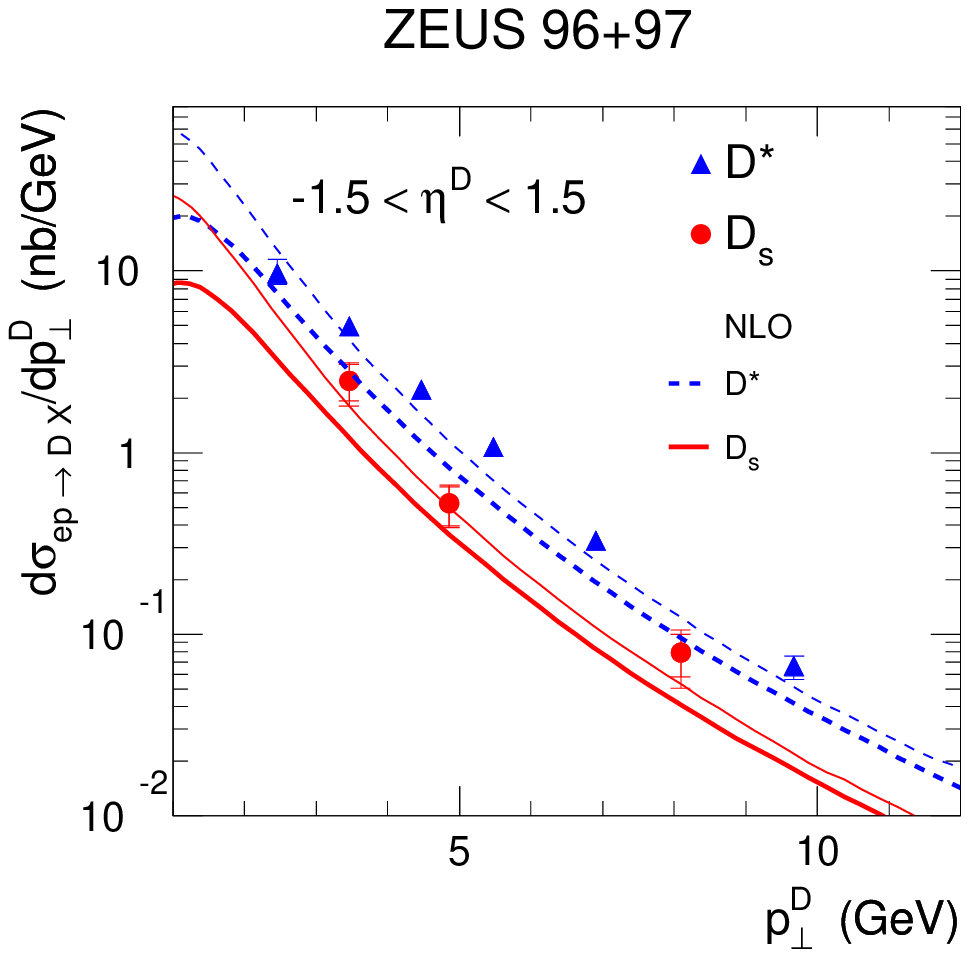,height=2.7in}
\epsfig{figure=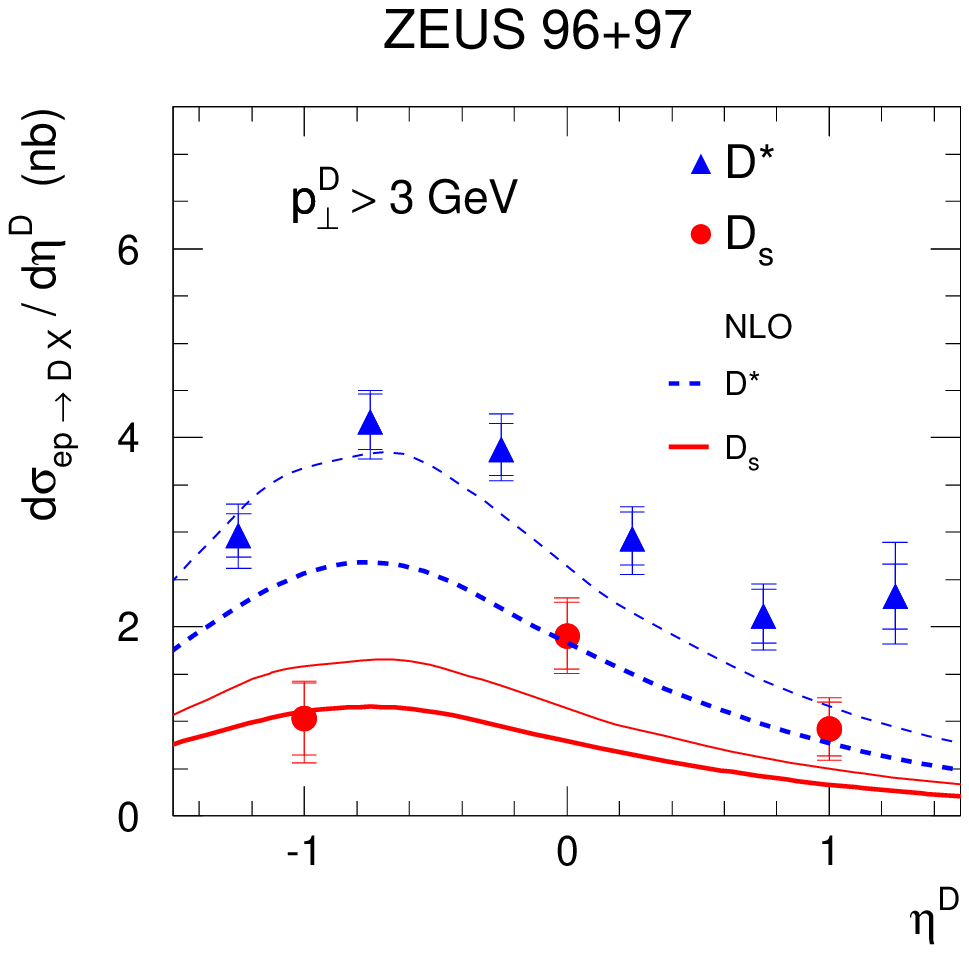,height=2.7in}
\caption{
Differential cross sections
$d\sigma/p_{\perp}^D$ (left) and $d\sigma/d\eta^D$ (right)
for \dsspm\ and \dspm\
photoproduction.
The \dsspm\ (dots) and \dspm\ (triangles) data
are compared with NLO predictions
for \dsspm\ (full curves) and \dspm\ (dashed curves).}
}

In Fig.~1, NLO calculations~\cite{MassC} obtained
with the MRS(G) and GRV-G~HO~\cite{pdf} parton density
parametrisations for the proton and photon, respectively,
are compared with the differential cross sections.
The thick curves were obtained with
the renormalization scale
$\mu_{R}=m_{\perp}\equiv\sqrt{ m_{c}^{2} + p_{\perp}^{2}} $ ($m_{c} =
1.5$\,GeV) and the factorization scales of the photon and proton
structure functions were set to $\mu_{F} = 2 m_{\perp}$.
For the thin curves, a rather extreme value for the pole $c$-quark mass,
$m_{c} =1.2$\,GeV, and a $\mu_{R}$ value of $0.5m_{\perp}$ were used.
The Peterson fragmentation function~\cite{PETER}
was used for charm fragmentation in this calculation.
Following the results of the NLO fits~\cite{nlofit}
to ARGUS data~\cite{argus},
the same values of the Peterson parameter, $\epsilon=0.035$, were used 
for both \dspm\ and \dsspm\ cross section calculations.
The fractions of $c$ quarks hadronizing as \dspm\ or \dsspm\
mesons were
extracted from results on charm production in
$e^+e^-$ annihilations~\cite{fractions}.
The NLO calculations underestimate the measured cross sections.
The shapes of the $p^D_{\perp}$ distributions are not completely reproduced.
For the $\eta^D$ distributions, the NLO predictions are below the data
in the central and forward (proton direction) regions.

An experimental separation of the direct and resolved processes was
obtained
by using the
variable $\xgo$, which is the fraction 
of the photon momentum contributing to the production of the two
jets with the highest transverse energies
within the accepted pseudorapidity range.
The charm photoproduction differential cross section was measured
as a function of  $\xgo$~\cite{dstar}.
A comparison of the $\xgo$
distribution with LO Monte Carlo (MC) simulations indicated the existence
of charm excitation in the photon parton density. Recently,
the ZEUS collaboration has performed new measurements of charm dijet
photoproduction~\cite{bud_499}.
The angle between the jet-jet axis and the beam axis in the dijet rest frame 
has been approximated by the variable $\cos\theta^*$, which is a function
of the pseudorapidities of the two jets:
$$\cos \theta^* = \tanh(\frac{\eta^{jet1}-\eta^{jet2}} {2}).$$

In the previous inclusive dijet analysis~\cite{dijet}, it was shown
that in direct photon processes, where the propagator in the 
leading-order (LO) QCD diagrams is a quark, the differential cross section 
$d\sigma /d|\cos\theta^* |$ rises slowly towards values of
$|\cos \theta^*| \sim 1$, while in resolved photon processes, where
in most cases                          the propagator
is        a gluon, it rises steeply with increasing
$|\cos \theta^*|$, consistent with the behavior of a spin-1 propagator.       \EPSFIGURE{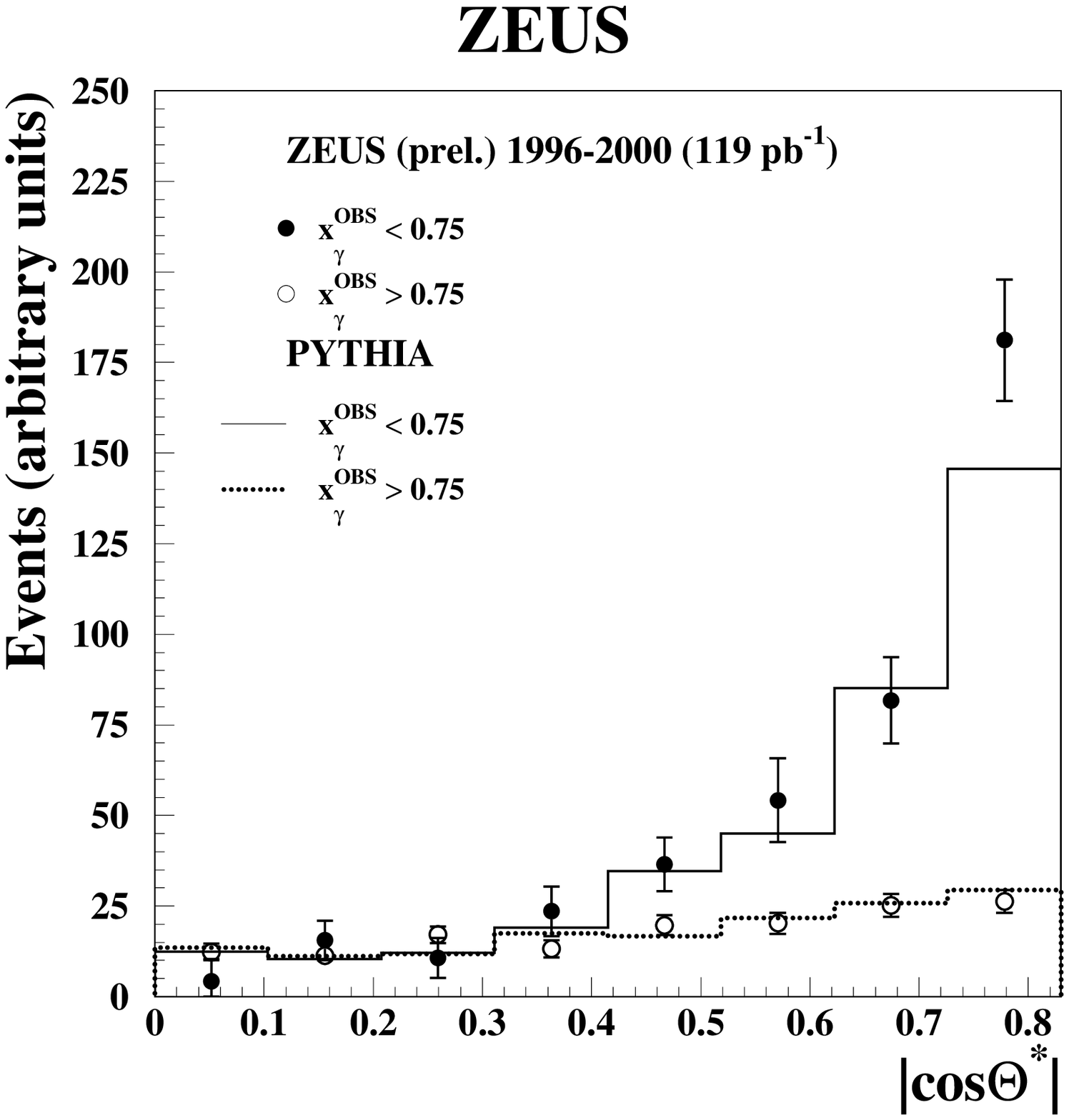,width=7.2cm}
{\label{fig:cos}
Differential distributions $dN /d|\cos\theta^*|$ for the data (dots)
and                         
PYTHIA MC simulation (lines).}
The differential distribution $dN /d|\cos\theta^*|$ for
events with at least two jets and a \dspm\ is shown in Fig.~2.
The data points are given separately for direct photon (open dots)
and for resolved photon (black dots) events.
In this analysis, a direct (resolved) photon process was
defined by the selection $\xgo>0.75$ ($\xgo<0.75$).
The dashed (full) histogram
is the PYTHIA~\cite{pythia} distribution for the direct (resolved) photon
MC events.
All the distributions are normalized to the resolved
data distribution in the first 4 bins.
An enhancement at high
values of $|\cos\theta^*|$ is seen for the resolved photon sample, both
in the data and in the MC simulation. The direct photon samples
do not show this strong peaking. This observation is consistent with
a significant gluon-exchange contribution and, consequently,
with a significant contribution of
charm excitation processes to charm photoproduction at HERA energies.

  \section{Beauty photoproduction}

Beauty photoproduction cross sections were measured by the H1
and ZEUS collaborations using events with at least two jets
and a lepton in the final state~\cite{h1_bb,zeus_bb}.
The beauty signals were extracted by fitting the $\ptr$
distributions of the data with Monte Carlo predictions for beauty
and the lighter flavor components, where $\ptr$ is the transverse
momentum of the lepton relative to the axis of the associated jet.
The cross sections were found to be
above NLO QCD expectations.

\EPSFIGURE{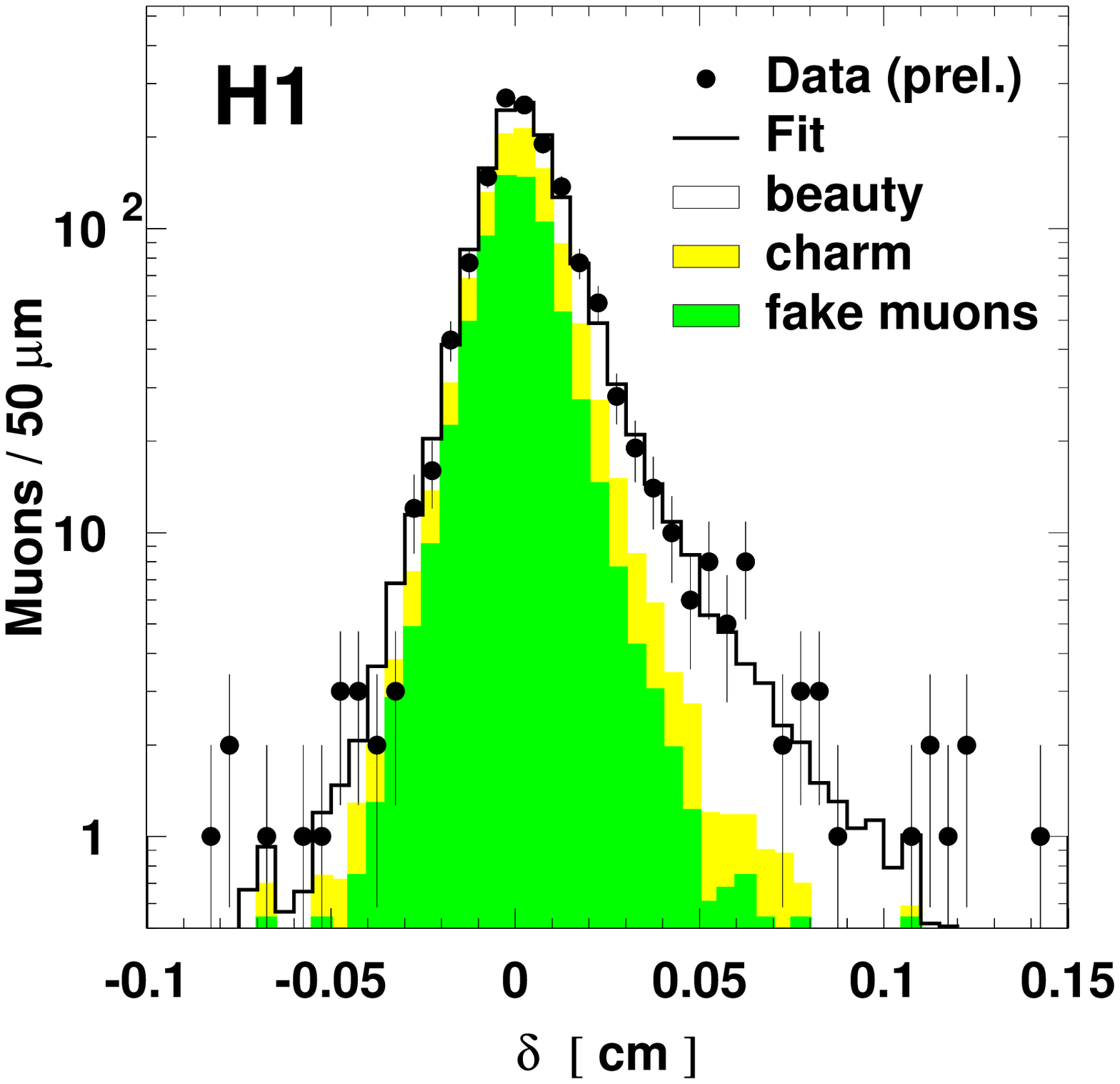,width=7.5cm}
{\label{fig:delta}
Muon impact parameter distribution and decomposition from the
likelihood fit.}
Recently, the H1 collaboration has performed new measurements
of the beauty photoproduction using a central silicon vertex
detector~\cite{osa_311}. 
The cross section was extracted from the $\ptr$ and impact
parameter ($\delta$) distributions of muons in dijet events.
Fig.~3 shows the observed impact parameter distribution
in the data together with histograms
indicating the contributions from beauty production and from
backgrounds. The decomposition was obtained from a likelihood fit
using the shapes of the $\delta$ distributions of beauty and charm events
from Monte Carlo simulations and of fake muons from real data.
Using both $\ptr$ and $\delta$ observables, and in combination
with the earlier result, the open beauty cross section was determined in
the visible range
$Q^2\,<\,1\,\rm{GeV}^2$, $0.1\,<\,y\,<\,0.8$, $p_{\perp}^{\mu}\,>\,2.0\,\rm{GeV}$,
$35^{\circ}\,<\,\theta^{\mu}\,< \,130^{\circ}$~:
$\sigma_{vis} (e p \rightarrow b \bar{b} X \rightarrow
   \mu X) = (170 \pm 25)\,{\rm pb}$.

The NLO QCD prediction, using the fixed-order NLO calculation~\cite{MassC}
and a fragmentation parameterization described in~\cite{h1_bb},
is $(54\pm 9)\,{\rm pb}$,
where the error was estimated by varying the renormalization
and factorization scales, the beauty quark mass value and
the fragmentation parameters.
The NLO prediction undershoots the measured cross section significantly.

The ZEUS collaboration has recently performed the first measurement
of the beauty photoproduction differential cross sections~\cite{bud_496}.
Events with a muon and at least two jets were selected.
Differential cross sections as a function of the muon pseudorapidity,
$\eta^\mu$, and transverse momentum, $p_T^\mu$, were calculated.
The beauty fraction in each bin of the selected variable was extracted
with the $\ptr$ fit.

\FIGURE{
\epsfig{figure=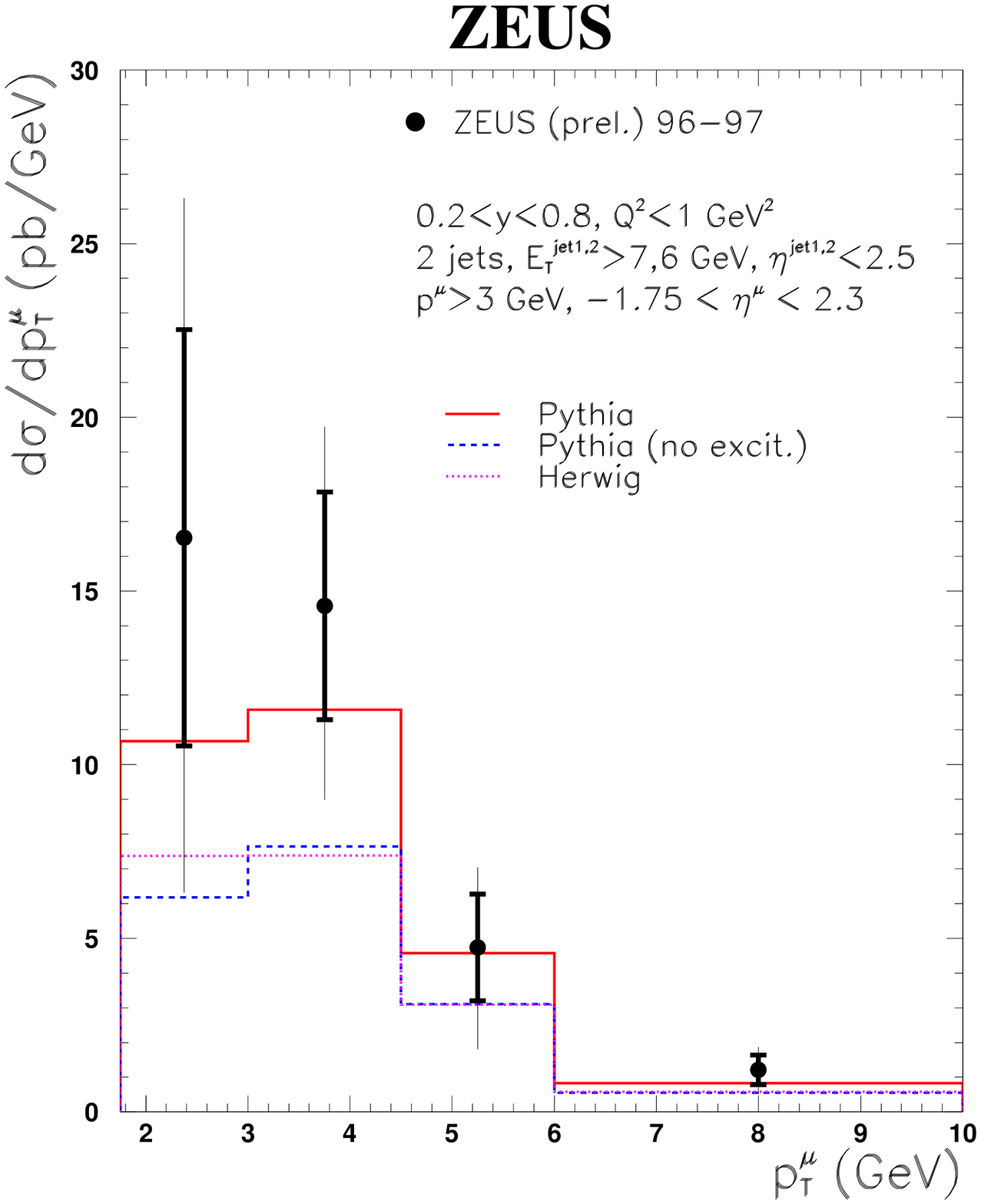,height=3.1in,width=2.9in}
\epsfig{figure=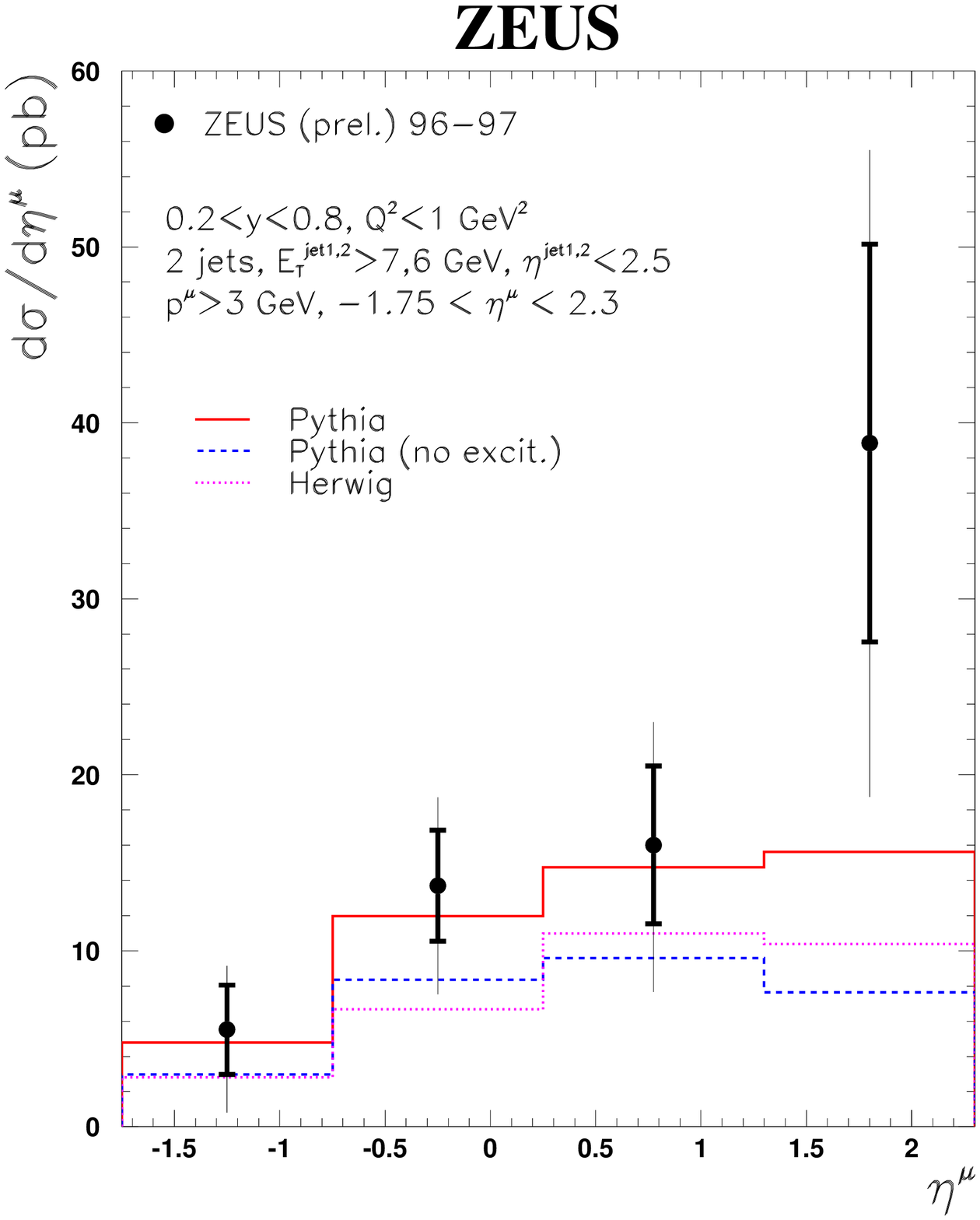,height=3.1in,width=2.9in}
\caption{
Differential beauty cross sections
$d\sigma/p_T^\mu$ (left) and $d\sigma/d\eta^\mu$ (right)
for events with two jets and a muon,
compared to predictions of PYTHIA and HERWIG MC programs.}
}
Fig.~4 shows the measured beauty differential cross sections in comparison
with predictions of PYTHIA~\cite{pythia} and HERWIG~\cite{herwig}.
The PYTHIA predictions are in reasonable agreement with the data.
The agreement is worse in the most-forward $\eta^\mu$ bin,
in which the contribution from the beauty excitation processes
is expected to be large. As shown, the beauty excitation is a substantial
component of the PYTHIA cross section. The HERWIG predictions, which also
include the beauty excitation component, are lower than PYTHIA
but still compatible with the data within errors.

  \section{Summary}

Charm and beauty photoproduction cross sections
have been measured by the H1 and ZEUS collaborations.
The perturbative QCD calculations are generally smaller than
the measured cross sections, in particular
in the forward (proton) direction.
In the charm sector, theoretical uncertainties
are larger than the experimental ones.
The study of dijet photoproduction associated with charm
is consistent with a significant contribution of charm excitation
processes.

\end{document}